\documentclass[12pt]{article}
\usepackage{graphicx} 

\newcommand{\ltap}{\
  \raise.3ex\hbox{$<$\kern-.75em\lower1ex\hbox{$\sim$}}\ } 
\newcommand{\gtap}{\
  \raise.3ex\hbox{$>$\kern-.75em\lower1ex\hbox{$\sim$}}\ }

\begin{document}

\title{A MATTER--ANTIMATTER UNIVERSE?}
\author{A.G. Cohen$^a$, A. De R\'ujula$^{b,a}$ and
  S.L. Glashow$^{c,a}$\ \thanks{\tt cohen@bu.edu,
    derujula@nxth21.cern.ch, glashow@physics.harvard.edu}\\ \\
\small \sl   $^a$Department of Physics, Boston University,
    Boston, MA 02215, USA \\
\small \sl   $^b$Theory Division, CERN,
 1211 Geneva 23, CH \\
\small \sl    $^c$Lyman Laboratory of Physics, Harvard University,
 Cambridge, MA 02138, USA \\
}


\date{}
\maketitle

\begin{abstract}  
We ask whether the universe can be a patchwork consisting of distinct
regions of matter and antimatter.  We demonstrate that, after
recombination, it is impossible to avoid annihilation near regional
boundaries.  We study the dynamics of this process to estimate two of
its signatures: a contribution to the cosmic diffuse $\gamma$-ray
background and a distortion of the cosmic microwave background. The
former signal exceeds observational limits unless the matter domain we
inhabit is virtually the entire visible universe. On general grounds,
we conclude that a matter--antimatter symmetric universe is
empirically excluded.
\end{abstract}

\section{Introduction and Outlook}
\label{sec:intro}

The laws of physics treat matter and antimatter almost symmetrically,
and yet the stars, dust and gas in our celestial neighborhood consist
exclusively of matter.  The absence of annihilation radiation from the
Virgo cluster shows that little antimatter is to be found within
$\sim\!20$~Mpc, the typical size of galactic clusters.  Furthermore,
its absence from X-ray-emitting clusters implies that these structures
do not contain significant admixtures of matter and antimatter.

Many cosmologists assume that the local dominance of matter over antimatter
persists throughout the entire visible universe. A vast literature
attempts to compute the baryonic asymmetry from first
principles. However, observational evidence for a {\em universal\/}
baryon asymmetry is weak.  In this regard, searches for
antimatter in cosmic radiation have been
proposed~\cite{antiprop}--\cite{AMSproposal}. Early next century, the 
AntiMatter Spectrometer (AMS), deployed aboard the International Space
Station Alpha~\cite{AMS,AMSproposal}, will search for antimatter in
space. Its reach is claimed to exceed 150~Mpc~\cite{APST}.  The
detection of cosmic anti-alpha particles would indicate the existence
of primordial antimatter; the detection of anti-nuclei with $Z>2$
would imply the existence of extra-galactic anti-stars.

The possible existence of distant deposits of cosmic antimatter has
been studied before \cite{Steigman}--\cite{Duda}. Steigman
\cite{Steigman} concluded that observations exclude significant
matter--antimatter admixtures in objects ranging in size from planets
to galactic clusters. Stecker {\it et al.} \cite{SMB} interpreted an
alleged shoulder in the cosmic diffuse gamma (CDG) spectrum near
1~MeV\footnote{We refer to cosmic diffuse photons by conventional
  names according to the current photon energy:
the night sky, the CDG and the CBR refer to visible,
$\sim 1$ MeV and  microwave photons, respectively.}
as relic $\gamma$-rays from antimatter annihilation. Recently,
Dudarewicz and Wolfendale \cite{Duda} used similar arguments to reach
a contrary conclusion: that the observed CDG spectrum rules out any
large antimatter domains.  These conflicting results are not based on
specific dynamics in a consistent cosmology. Our analysis uses current
data and avoids {\it ad hoc\/} assumptions concerning a
matter--antimatter universe.

We explore the possibility of universal (but not local)
matter-antimatter symmetry.  In what we term the $B=0$ universe, space
is divided into regions populated exclusively by matter or
antimatter. Our conclusions do not depend on how this structure
evolved, but it is reassuring to have an explicit model in mind:
consider an inflationary cosmology in which baryon (or antibaryon)
excesses develop in the manner suggested by Sakharov
\cite{Sakharov}. In models with spontaneous CP violation, the
Lagrangian may be chosen judiciously so that the `sign' of CP
violation (determining whether a local baryon or antibaryon excess
develops) is randomly and abruptly assigned to regions as they emerge
from their horizons during inflation. Soon after baryogenesis, the
domain walls separating matter and antimatter evaporate.  As regions
of matter or antimatter later re-enter their horizons, the $B=0$
universe becomes a two-phase distribution.

Let today's domains be characterized by a size $d_0$, such that
$1/d_0$ is their mean surface-to-volume ratio.  Because the existence
of anti-galaxies within a matter-dominated domain is empirically
excluded, we must (and can) arrange the distribution of domains to be
sharply cut off at sizes smaller than $d_0$.  Explicit inflationary
models satisfying these constraints exist \cite{CDG}, but are
described no further because we find all such models to conflict with
observation.

The current domain size $d_0$ is the only parameter of the $B=0$
universe crucial to the confrontation of theory with observation. To
agree with constraints from X-ray-emitting clusters, $d_0$ must exceed
a minimal value, $\sim 20$~Mpc.  For $d_0=20$~Mpc, the
visible universe would consist of $\sim\! 10^7$ domains.  We derive a
stronger lower limit on $d_0$ comparable to the current size of the
visible universe, thereby excluding the $B=0$ universe.

An explicit cosmological model is necessary to estimate the observable
signals produced by annihilation. We assume a Robertson--Walker
universe and use fiducial values for the relevant cosmological
parameters: critical mass density $\Omega =1$; vanishing cosmological
constant $\Omega_\Lambda=0$; Hubble constant $H_0=75$~km/s$\cdot$Mpc
or $h=0.75$; and an average baryon (or antibaryon) number density
$n_B\equiv \eta \, n_\gamma$ with $\eta=2\times 10^{-10}$.

In Section~\ref{sec:cosmo} we
show that our conclusions are unaffected by other choices for
$\Omega$, $\Omega_\Lambda$ and $H_0$ within their empirically allowed
domains.  Consequently we do not express our results explicitly in
terms of these cosmological parameters.
The annihilation signals we study depend linearly on $\eta$. To
compute lower limits to the signals, we chose $\eta$ at the low end of
the domain allowed by analyses of primordial element abundances. (For
a recent discussion, see \cite{HSBL}.)  

In Section~\ref{sec:era} we explain why particle--antiparticle
annihilation is unavoidable from the time of recombination to the
onset of structure formation.  Following a conservative approach, we
consider only those annihilations occurring during this period.  Our
analysis involves known principles of particle, atomic and plasma
physics, but the dynamics of the annihilating fluids (discussed in
Section~\ref{sec:encounter}) is complicated and the considerations
required to reach our results are elaborate.

The immediate products of nuclear annihilation are primarily pions
($\pi^+$, $\pi^0$ and $\pi^-$) with similar multiplicities and energy
spectra. The end products are energetic photons from $\pi^0$ decay,
energetic electrons\footnote{Relativistic electrons and positrons behave 
similarly, and we refer to both as electrons.}  ($e^+$ and
$e^-$) from the decay chain $\pi\rightarrow \mu\rightarrow e$, and
neutrinos. Although they are produced at cosmological distances, the
annihilation photons and electrons  can each
produce potentially observable signals:

\begin{itemize}  
\item{} The energy carried off by annihilation electrons (about
  320~MeV per annihilation) affects the CBR spectrum directly (via
  Compton scattering) and indirectly (by heating the medium). The
  consequent distortion of the CBR (discussed in
  Section~\ref{sec:distort}) cannot exceed observational limits.
\item{} Most of the annihilation photons, although redshifted,
  are still present in the universe. Their flux
  (computed in Section~\ref{sec:diffuse}) cannot exceed the observed
CDG flux. 
\end{itemize}
Because these signals increase inversely with the domain size,
our analysis yields a lower limit for $d_0$.
In fact, we obtain no new constraint from comparing the expected
distortion of the CBR with its measured limits.  However, the CDG flux
produced by annihilation far exceeds the observed flux unless $d_0$
is comparable in size to the visible universe.
Thus, the  $B=0$ universe is
excluded. 

\section{The Era of Unavoidable Annihilation}
\label{sec:era}

What if the matter and antimatter domains are and have always been
spatially separated?  If large empty voids lay between them, there would be
no observable annihilation signals. We now show how the observed uniformity
of the CBR rules voids out.

Two events took place at roughly the same time in cosmic
history: the transition from charged plasma to neutral atoms
(recombination) and the decoupling of radiation and ordinary matter
(last scattering). For our fiducial cosmological parameters, these
events occurred at a temperature $\sim\!0.25$~eV and at a redshift
$y_R\simeq1100$ (we use $y\equiv 1+z = 1/R(t)$ as a redshift parameter, rather
than the conventional $z$). The transition to transparency was not
instantaneous, but evolved during an interval $y_R\pm 100$ whose
half-width is $\sim\!15$~Mpc in comoving (current)
distance units. Thus, features at recombination of comoving size
smaller than $15$~Mpc cannot be discerned in the CBR.

Large-scale non-uniformities of the matter density, whether dark or
baryonic \cite{SW} generate variations of the CBR temperature. Its observed
uniformity (to parts in $10^{-5}$) implies a very uniform  density of 
ordinary matter at $y=y_R\simeq1100$, to within the resolution discussed
above. It follows that  voids between matter and
antimatter domains must be smaller than $15$~Mpc.

The baryon density depletion in voids is damped as photons diffuse
toward  less dense regions, dragging matter with them. By
recombination, inhomogeneities with current size $\ltap\! 16$~Mpc would be
destroyed\footnote{This assumes that 
  such inhomogeneities are not strictly isothermal, a situation
  considered in Section~\ref{sec:cosmo}.} by this mechanism
\cite{Silk}. This upper bound coincides with 
the smallest resolvable structure in the CBR.  Thus, voids large enough to
survive until recombination would have been detected. While matter
and antimatter regions may have been separated prior to recombination, they
must be in immediate contact afterward. 
Thus, in determining the minimal signal of a $B=0$ universe, we do not
consider annihilations occurring at $y>y_R$.

The mechanism by which the nearly uniform universe at large $y$
evolved today's large-scale structures is not well understood. We
cannot confidently assert what effects this will have on annihilation
in a $B=0$ universe. It
could well be
that the collapse of baryonic matter into galaxies and stars
quenches annihilation unless the collapsing system overlaps a domain
boundary, a situation we consider shortly.  Our conservative estimate of
the annihilation signal includes matter--antimatter annihilation taking
place prior to the redshift $y_S$ at which the earliest density fluctuations
become large ($\delta \rho / \rho \sim 1$). We take $y_S\simeq 20$, which
is estimated to be the epoch of galactic condensation \cite{Peebles2} and
earliest star formation \cite{Peebles3}.  We compute the 
signals due to annihilations taking place during the interval $1100 > y > 20$.

The large-scale density contrast of the visible universe need not
coincide with  
the pattern of matter and antimatter domains. 
A density fluctuation beginning to collapse could overlap a
domain boundary. Successful collapse would
yield a structure with a significant mixture of matter and antimatter. In
this case  annihilation would proceed even more rapidly at the onset of
structure formation. Yet we cannot be confident that such mixed
structures form.

In the linear regime ($\delta \rho / \rho \ll 1$), the mean annihilation
rate is not affected by density fluctuations. But, what happens as the
fluctuations grow? If an over-density is to overcome expansion and become a
self-gravitating system, it must satisfy the Jeans condition: the sound
travel time across the object $l/v_s$ must be greater than the
characteristic free-fall time $1/\sqrt{G\, \rho}$, or $G\, l^2 \, \rho \ge
v_s^2$. Suppose that equality is approached by an over-density containing both
matter and antimatter. Further contraction increases the annihilation rate,
thus reducing $\rho$ and driving the system away from collapse.  Thus,
our conservative estimate of the annihilation signal assumes that
density fluctuations straddling  domain boundaries either fail to collapse
or form separate unmixed structures.

\section{The Matter--Antimatter Encounter}
\label{sec:encounter}

The $B=0$ universe consists of matter and antimatter domains with almost
identical mass densities that, as we have shown, 
touch one another from recombination to the 
onset of structure formation. As annihilation proceeds near an interface, a
flow develops as new fluid replenishes what is annihilated. This  flow must
be analyzed to determine the annihilation rate on which our putative
signals depend.  The analysis involves established and well-understood
principles of physics, but is complicated by the energy released by nuclear
annihilation.  (We neglect $e^+ e^-$ annihilation, whose energy release is
much smaller.)  The processes by which annihilation electrons lose
energy produce crucial effects on the ambient fluid, as well as a
potentially observable distortion of the CBR. (High-energy photons
from $\pi^0$ decay, although  responsible for the CDG signal, have little
effect on the medium through which they pass.)

The primary energy-loss mechanism of the annihilation electrons  is 
Compton scattering off CBR photons (see Appendix~\ref{app:electrons}).
This process up-scatters target photons to higher energies. The resultant 
flux of UV photons heats and ionizes ambient matter throughout much of
the universe and for all of the relevant period.  Moreover, the annihilation
electrons lose  a small portion of their initial energies by scattering off
ambient electrons in the fluid. This process heats the fluid within the
electron range, thereby  accelerating the flow and leading to
even more annihilation---a feedback mechanism making the matter--antimatter
encounter potentially explosive.

\begin{figure}[htb]
  \centering
  \includegraphics[angle=-90,width=\textwidth,]{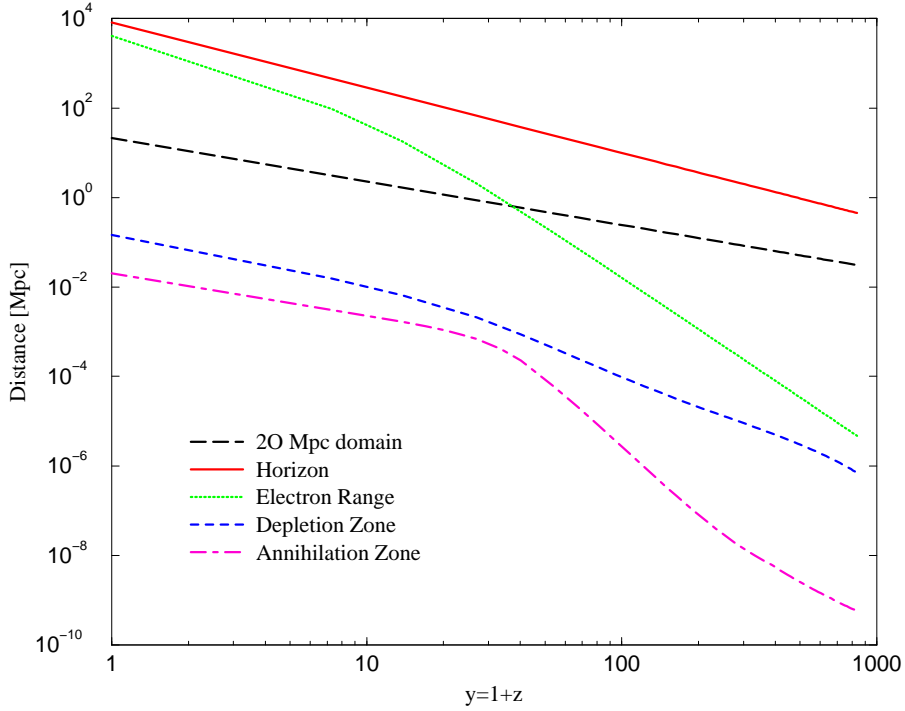}
  \caption{Read from the top at large $y$: the horizon, the look-back size of a
  20 Mpc domain, the widths of the reheated zone, the depletion zone,
and the annihilation zone.} 
\label{fig:sizes}
\end{figure}

Several length scales characterize the fluid dynamics about
a matter--antimatter interface. They are:
\begin{itemize} 
\item{} $A$,  the width of the {\it annihilation zone,} wherein both
  matter and antimatter are present;
\item{} $D$, the width of the {\it depletion zone,} wherein fluid
  flow toward the annihilation zone reduces the density;
\item{} $L$, the width of the {\it reheated zone,} wherein electrons
  produced by annihilations directly deposit energy into the
  fluid. This is simply the electron range.
\end{itemize}
These length scales, computed in Appendix~\ref{app:range} and
later in this section, are shown in Fig.~\ref{fig:sizes}
along with 
a comoving domain size of $20$~Mpc and the horizon scale. Annihilation
takes place in the vicinity of the domain boundary and well within
the depletion zone, which itself is much shorter than the electron
range. That is, in the relevant redshift domain: $A\ll D\ll L$. 
This distance
hierarchy lets us treat the flow as one-dimensional.

Annihilation has a negligible effect on the CBR temperature
$T_\gamma(y)$, which remains 
as it is in a conventional universe. However, the annihilation debris
produce and maintain virtually total ionization,
as shown in Appendix~\ref{app:ions}.
Therefore the annihilating fluid consists of
photons, protons, antiprotons, electrons and
positrons\footnote{We neglect the helium contamination ($\sim\! 7\,\%$
  by number) and those of larger primordial nuclei.}.
The proton and electron number densities coincide,
except in the narrow annihilation zone.
Consequently our analysis may be put in terms of
the total matter mass
density $\rho \equiv m_e n_e + m_p n_p$, the total fluid momentum density
$\rho v$, the
total fluid pressure $p$ and the total fluid energy density $\epsilon
= \epsilon_{\rm thermal} + \rho v^2/2$.
The internal energy density and pressure are related as for a
non-relativistic ideal gas: $\epsilon_{\rm thermal} = 3\,p/2$.

The equations describing the flow of matter are 
conservation laws
for particle number (mass in the non-relativistic limit), momentum, and
energy. They must take account of the following phenomena:

\begin{itemize}
\item The depletion of fluid mass, momentum and energy by
  nuclear annihilation. 
\item The effect of the CBR on the fluid momentum
  and energy. 
\item The effect of the annihilation products on the
  fluid  momentum and energy.  
\item The expansion of the universe.
\end{itemize}
The expansion of the universe is taken into account
by expressing the conservation laws in a Roberston--Walker
universe~\cite{Weinberg}. 
The metric is $ds^2=dt^2-R^2(t)d\chi^2$, with $\chi$ a comoving
spatial coordinate normal to a domain boundary.
The remaining effects are dealt with by including
appropriate source terms in the fluid equations:

\begin{equation}
  \label{density}
  \frac{\partial \rho}{\partial t} + 3 \,\frac{\dot R}{R}\, \rho +
  \frac{1}{R}\frac{\partial (\rho v)}{\partial
    \chi} = -\Gamma_{\rm ann}\, \rho
\end{equation}
\begin{equation}
  \label{momentum}
  \frac{\partial (\rho v)}{\partial t} + 4\, \frac{\dot R}{R}\, \rho v +
  \frac{1}{R}\frac{\partial}{\partial \chi}
  \bigl[\rho v^2 + p \bigr]
  = -\Gamma_{\rm ann}\, \rho v - \frac{4}{3}\, \sigma_T \,\frac{
    u_\gamma}{m_p c} \, \rho v 
\end{equation}
\begin{equation}
  \label{energy}
  \frac{\partial \epsilon}{\partial t} + 5\, \frac{\dot R}{R}\, \epsilon +
  \frac{1}{R}\frac{\partial}{\partial \chi}
  \bigl[(\epsilon + p) v \bigr]
  = -\Gamma_{\rm ann} \epsilon - \frac{4}{3}\, \sigma_T\, \frac{
    u_\gamma}{m_e c} \left(\frac{3\,p}{2} -  \frac{3\,\rho\, T_\gamma}{m_p}
  \right) + H_\epsilon\;.
\end{equation}
Here $\Gamma_{\rm ann}\equiv\langle \sigma_{\rm ann} v\rangle \bar n_p$
is the matter annihilation rate, 
$\sigma_T$ is the Thompson cross section and $u_\gamma$ is the
CBR energy density.
The terms involving $\sigma_T$ describe
the transfer of energy and momentum between  the fluid and the CBR
resulting from Compton scattering. 
$H_\epsilon$, given by
Eq.~(\ref{Heat}) in Appendix~\ref{app:electrons},
is the rate of change of the
energy density of the fluid due to its interactions with the
annihilation debris. It receives a direct
contribution from the annihilation electrons,
and an indirect one from UV photons up-scattered by Compton collisions of
these electrons with the CBR. We find that
the contribution of the electrons dominates
within the electron range. Beyond this range, only the UV photons 
contribute to $H_\epsilon$.
In Eq.~(\ref{momentum}), we have neglected the small
contribution by the annihilation debris to the fluid momentum. 

The signals of a $B=0$ universe---the CDG and a distortion of the
CBR---are functions of $J$,
the number of annihilations taking place per unit
time and area orthogonal to the surface of an annihilation zone:
\begin{equation}
 J\equiv \int \langle \sigma_{\rm ann} v_{p\bar p} \rangle\, n_p
  \bar n_p \,R\,d\chi \ ,
  \label{annrate}
\end{equation}
where the integral extends over a
single annihilation zone  with $\chi = 0$ at
its mid-point, and $\bar n_p(\chi) = n_p(-\chi)$. 
The width of the annihilation zone $A$ may be estimated as
  $A\sim J/\langle \sigma_{p\bar p} v\rangle n_\infty^2$, where~\cite{MoHu}
  $\sigma_{p\bar p} v \simeq 6.5 \times 10^{-17}\; {\rm cm}^3\,
  {\rm s}^{-1}\,{c/v}$ and $n_\infty$ is the proton density far from
  the annihilation zone.
We must solve Eqs.~(\ref{density}--\ref{energy}) to determine $J$.

\subsection{A Qualitative Solution}

Because our fluid equations do not admit analytic solutions we begin
with a qualitative discussion.
The value of $\Gamma_{\rm ann}$ is always much greater than the
expansion rate, so that the solutions to
Eqs.~(\ref{density})--(\ref{energy}) rapidly reach 
equilibrium in the annihilation zone. Consequently, taking the
limit $\sigma_{\rm ann} \to\infty$  yields a good approximation.
In this limit, the width of the
annihilation zone shrinks to zero and the annihilation terms in the
fluid equations may be replaced by a boundary condition at the domain
interface. The rate of annihilation per unit
surface area is then given by the proton flux at the interface:
\begin{equation}
 J \simeq \frac{\rho(\chi,t)}{m_p}\, v(\chi,t)\bigg|_{\chi=0} \ .
  \label{annrate2}
\end{equation}

Two effects result from 
the couplings of the fluid to the CBR. 
The term in
Eq.~(\ref{momentum}) proportional to $\sigma_T$
tends to damp 
the fluid motion. The corresponding term in
Eq.~(\ref{energy}) tends to keep the fluid temperature near $T_\gamma$. 
For $y \gtap 400$ these terms dominate, so that the
two temperatures are locked together, $T\simeq T_\gamma$. The CBR drag 
on the fluid leads to diffusive motion, and we may define
a time-dependent diffusion constant:
\begin{equation}
D_{e \gamma}\equiv{45\over 4\, \pi^2\, \sigma_T\, T_\gamma^3}\,.
\label{Diffcon}
\end{equation}
The solution to the resulting diffusion-like equations 
gives an estimate of the annihilation rate $J$:
\begin{equation}
J\simeq n_\infty (t) \, \sqrt{{5\,D_{e \gamma}\over 3\, \pi \, t}}
\label{diffchapuza}
\end{equation}
with $n_\infty$ the proton number density far from the interface.
The width of the depletion zone is comparable to the diffusion length
$D\sim \sqrt{D_{e\gamma}\, t}$.

For redshifts $y \ltap 200$ the effects of the CBR on the
fluid motion are negligible and we may ignore terms
proportional to $\sigma_T$.  In
this case,  which we refer to as `hydrodynamic',  the motion is
controlled by pressure gradients and the fluid flows at
a substantial fraction of the speed of
sound.
The resulting equations are those describing a  gas
expanding into a semi-infinite vacuum in the
presence of an energy source. 
An analytic solution exists for $H_\epsilon=0$. In this case 
the annihilation rate  $J$ is
\begin{equation}
J=
C\; n_\infty (t)\; v_\infty(t)=
C\;n_\infty(t)\;\sqrt{5\,T_\infty(t)\over 3\, m_p}
\label{chapuza}
\end{equation}
with $v_\infty$ the speed of sound 
and $T_\infty$ the fluid temperature $T=p/(n_p+n_e)$ far from the
annihilation zone. 
The coefficient of proportionality is\footnote{This is the
  adiabatic solution. For $100 \ltap
  y\ltap 200$ the process is more nearly isothermal. The corresponding
  value of $C$ is $1/e$.} $C=(3/4)^4$. 
The width of the depletion zone in this case is comparable to the
sound-travel distance $D\sim R(t) \int^t dt'\,v_\infty/R$.

In the intermediate region, $200 \ltap y \ltap 400$, neither of the
above approximations give a quantitatively accurate picture of the
fluid motion.

\subsection{The Numerical Solution}

We have integrated Eqs.~(\ref{density}--\ref{energy}) numerically
to determine the fluid temperature $T$ and the annihilation rate
per unit surface area $J$ near a domain boundary.
The diffusive nature of the solution at large $y$
has a welcome consequence: all memory of the initial
conditions is lost as the fluid evolves. The post-recombination
annihilation signal does not depend on the (pre-recombination) time at
which matter and antimatter domains first come into contact. To solve
Eqs.~(\ref{density})--(\ref{energy})
we 
choose initial conditions at recombination such that the
matter and antimatter domains have
constant density, have no peculiar velocity and
touch along the surface $\chi=0$.
Our results are more conveniently presented in terms of $y$
rather than time, according to $dy=-y\,H(y)\, dt$.
For our fiducial choice of cosmological
parameters $H(y)=H_0\,y^{3/2}$.

\begin{figure}[htb]
  \centering \includegraphics[angle=-90,width=\textwidth,]{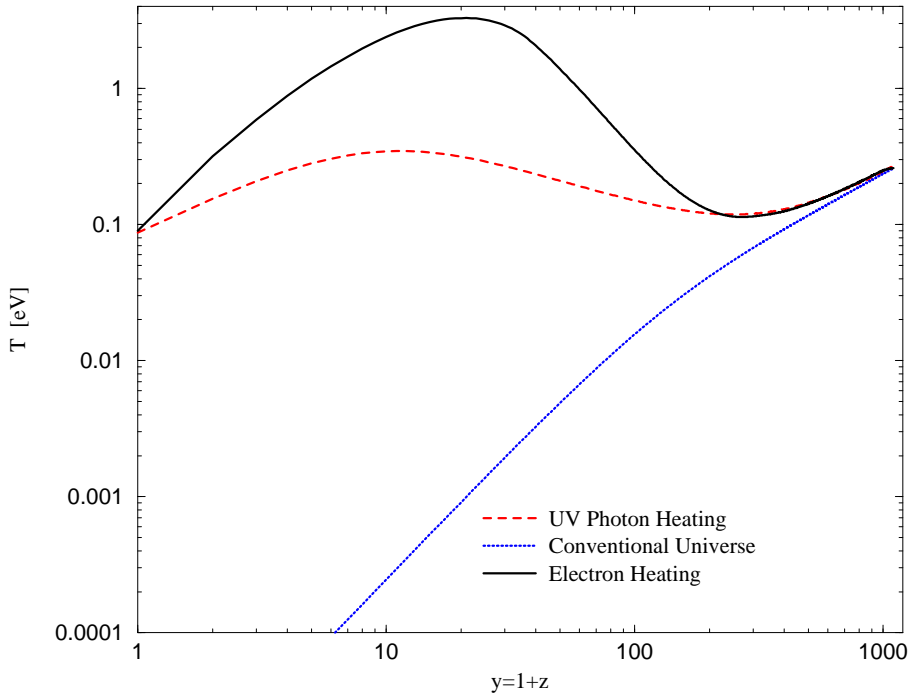}
  \caption{Temperatures (in eV) as  functions of redshift
    $y\equiv 1+z$ .} 
\label{fig:temperature}
\end{figure}

The dotted curve in Fig.~\ref{fig:temperature} is
the fluid temperature $T(y)$  
in a conventional universe.
The remaining two curves are the computed temperatures of the $B=0$ universe:
the solid curve is $T(y)$ within the electron range where
heating by relativistic electrons dominates; the dashed curve
is its value outside this region, where UV photons are the only heat source.
For 
$y \gtap 400$,  the 
CBR is an effective heat bath keeping matter and radiation close to thermal
equilibrium. Heating due
to the annihilation products plays an important role at lower $y$:
it increases the  fluid temperature leading to a larger fluid velocity.
According to 
Eq.~(\ref{annrate2}) the annihilation rate $J$ is thereby enhanced.

\begin{figure}[htb]
  \centering \includegraphics[angle=-90,width=\textwidth,]{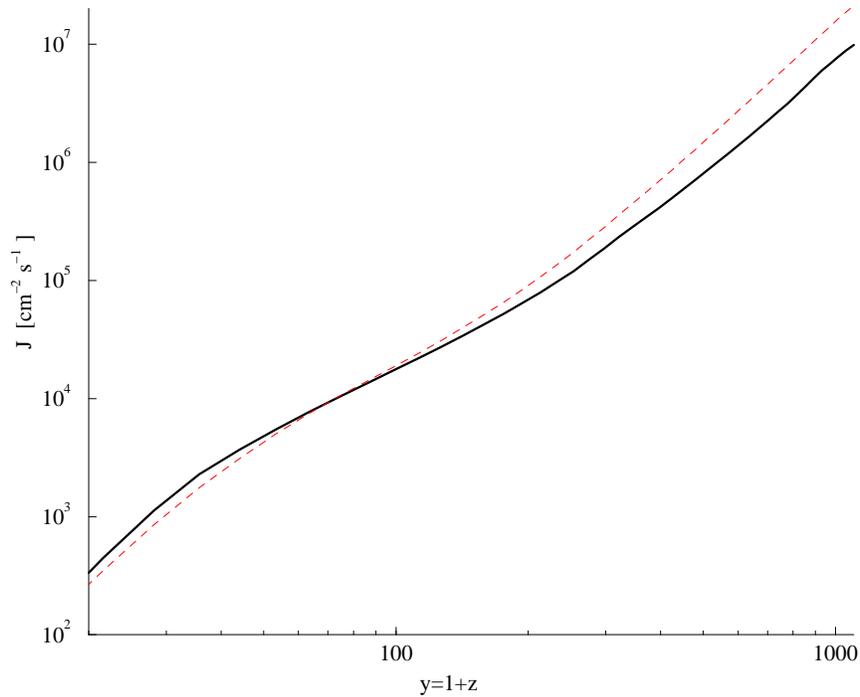}
  \caption{Annihilation rate $J$ (in particles per
    cm$^2$ per second) as a function of redshift $y$.  The solid curve is
    our numerical solution; the dashed curve is an approximate
    result discussed in the text.} 
\label{fig:current}
\end{figure}

The solid curve in Fig.~\ref{fig:current} is our numerical
result for $J$, the annihilation rate per unit surface area defined
by Eq.~(\ref{annrate}). The 
dashed curve is the approximation given by 
Eq.~(\ref{chapuza}) using the temperature obtained from the
numerical integration.
Although  its derivation ignored $H_\epsilon$,
Eq.~(\ref{chapuza}) 
agrees quite well with our numerical result for this choice of $T(y)$.
At larger
redshifts, the motion is diffusive and our
numerical result should be (and is)
substantially less than the qualitative hydrodynamic
estimate\footnote{Our diffusive results for $y\gtap 400$ are at
  variance with those of~\cite{kandt}, where the annihilation rate
  prior to recombination is estimated on the basis of proton
  free-streaming.}, as is seen in the figure. 
Had we used the matter temperature of a conventional 
universe in Eq.~(\ref{chapuza}), we would have obtained an annihilation 
rate nearly two orders of magnitude smaller. 
The heating of the fluid by annihilation debris (described by $H_\epsilon$)
has a dramatic effect on the annihilation rate
$J$, and {\it a fortiori\/} on the consequent signals of the $B=0$
universe. 

At all redshifts the annihilation rate is determined
by the flow of the highly ionized matter and antimatter 
fluids into the annihilation zone.
The momentum-transfer cross section
$\sigma_C$ in proton-antiproton Coulomb collisions, which controls
diffusive mixing of these fluids, is large compared to the
annihilation cross section 
$\sigma_{\rm ann}$.  If mixing results only from diffusion, as 
in quasi-static laminar flow, the annihilation current would be reduced
by a factor $\sim (\sigma_{\rm ann}/\sigma_C)^{1/2}$ relative to $J$.
However turbulence produces full mixing in the annihilation zone,
while leaving the average flow unaffected, thus justifying our neglect
of the Coulomb scattering term in equations (1--3).

An analysis of fluctuations about a laminar flow into the
annihilation zone demonstrates an instability towards turbulent
mixing (this is analogous to the instability of a planar flame front
in combustive flow~\cite{landau}.)  For the width $A$ of
the annihilation zone we have obtained, the Reynolds number is
${\cal{R}}_A \sim A\,\sigma_C\,n > 10^5$, large enough to insure a
turbulent flow at this and larger scales.
Turbulence efficiently mixes the fluids in the
annihilation zone, but does not significantly retard their mean motion
in the depletion zone. Thus turbulence drives the flow to the
solution we have discussed, wherein the annihilation rate is
determined solely by the rate at which material can be transported
toward the annihilation zone.

The drag on the fluid exerted by the CBR and the velocity redshift due
to expansion suppress turbulence on large scales.  The first effect
dominates during the redshift range of interest, suppressing turbulence
for scales $\lambda > v \, m_p\,c /(\sigma_T\, u_\gamma)$, which is
larger than the width $A$ of the annihilation domain.

\section{Distortion of the CBR}
\label{sec:distort}

Measurements of the CBR, being much more precise than those of the
CDG, might be expected to provide the most stringent constraint on the
$B=0$ universe.  In this section, we use our conservative calculation
of the annihilation rate to estimate the distortion of the CBR
spectrum.  In performing this calculation, we make several
approximations that somewhat overestimate the effect. Nonetheless,
the consequent distortion lies well below the observed limit, and
provides no constraint at all.

Annihilation produces relativistic electrons  and energetic photons.  
Annihilation electrons have a direct effect on the CBR by scattering
photons to higher energies, thereby skewing the CBR spectrum. Moreover
these electrons heat the ambient plasma. The heated plasma
produces an additional indirect spectral distortion.
(The energetic photons from neutral pion decay have energies too high
to have much effect on the cosmic microwave background.)

To compute the direct effect, we must determine the
number  of  CBR photons
scattered from energy $\omega_i$ to $\omega_f$ by a single 
electron. This function, $d^2N(\omega_f, \omega_i)/d\omega_f\,
d\omega_i$, is computed in Appendix~\ref{app:cbr}. 
The electron multiplicity per $p\bar p$ annihilation is similar to the 
photon multiplicity, measured~\cite{Rgamma} to be $\bar g \simeq 3.8$.
The number of annihilation electrons made per
unit volume and time is $ {\bar g}\,{J}/d$,
where $1/d\equiv y/d_0$ is the average domain surface-to-volume ratio at
epoch $y$. The
spectral distortion $\delta u_\gamma(\omega)$ (energy per unit volume
and energy) satisfies a transport equation:
\begin{eqnarray}
  \left( y\,{\partial\over \partial y}
    +\omega\, {\partial\over \partial \omega} -3 \right)\,\delta
  u_\gamma(\omega,y) =& \nonumber\\
  {\omega {\bar g}\, J(y)\over H(y)\, d(y)} \int d\nu\left(\frac{d^2 
N(\nu, \omega)}{d\nu
    d\omega} - 
  \frac{d^2 N(\omega, \nu)}{d\nu d\omega}\right)
  &\equiv A(\omega,y)\;.
\label{deltau}
\end{eqnarray}
We have ignored absorption of UV photons by neutral hydrogen because
the $B=0$ universe is largely ionized.

\begin{figure}[htb]
  \centering \includegraphics[angle=-90,width=\textwidth]{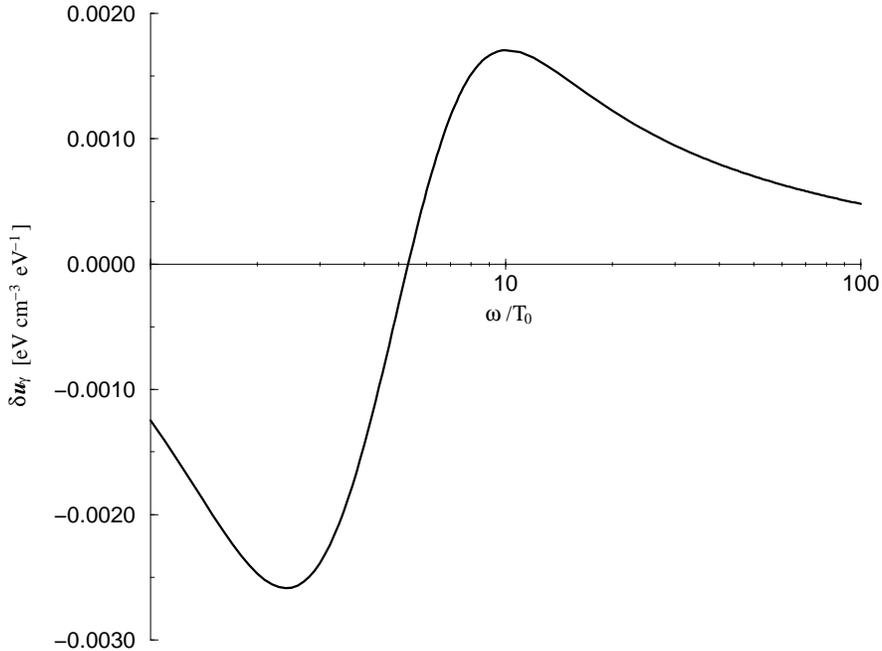}
  \caption{The CBR spectral distortion.
    Beyond the range shown, $\delta u_\gamma \propto 1/\sqrt{\omega}$,
    up to $\omega/T_0 \sim 10^4$.} 
  \label{fig:deltaugamma}
\end{figure}

The direct contribution to the CBR distortion is the
 solution to Eq.~(\ref{deltau}) evaluated at the current epoch:
$\delta u_\gamma(\omega)\equiv \delta u_\gamma(\omega,1)$.
It is given by:
\begin{equation}
  \delta u_\gamma(\omega) =
  \int^{y_S}_{y_R} \frac{dy}{y^4}\, A(\omega\, 
    y,y)\ ,
\end{equation}
where we have confined the source to $1100>y>20$, the era
of unavoidable annihilation. 
To evaluate the integral we use the
annihilation rate $J$ computed in Section~\ref{sec:encounter}. 
Figure~\ref{fig:deltaugamma} displays the result
for a  current domain size of $20$ Mpc.
Note that
$\vert\delta u_\gamma(\omega)\vert$ is always less than 
$3\times10^{-3}$~cm$^{-3}\simeq 1.8\times10^{-6}\;T_0^3$.
The limit set by COBE--FIRAS ~\cite{COBE} 
on  rms departures from a thermal spectrum is
$\vert \delta u_\gamma(\omega) \vert < 7.2\times 10^{-6}\, T_0^3$
throughout the energy range $T_0 < \omega < 10\;T_0$.
This upper limit is four times larger than our computed signal for
the minimum domain size. Because
larger domains yield proportionally smaller results,
we obtain no constraint on the $B=0$ universe.

The indirect contribution to the
CBR distortion results from a temperature difference $T-T_\gamma$
between the heated ambient fluid and the CBR. It may be described by the
Sunyaev--Zeldovich parameter $Y$ \cite{suny}:
\begin{equation}
Y  =  \int
\frac{\sigma_T \, n_e (T-T_\gamma)}{m_e c^2}\, dl\;,
\end{equation}
where the integral is along the photon path $dl=-c\,dy/y\,H(y)$. 

Within the
electron range, collisions between 
annihilation electrons and the plasma
 result in a temperature profile $T(y)$ shown as the solid curve in
Fig.~\ref{fig:temperature}. Outside the electron range, reheating
is due to photons up-scattered by these electrons,
resulting in the 
temperature profile shown as the dashed curve. 
CBR photons 
may have traversed regions of both types. 
To compute $Y$, we use
the higher temperature profile
(the one within the electron range). We  thereby 
overestimate the signal. Our
result is $Y \ltap 9 \times10^{-7}$, which 
is over an order of magnitude below  the COBE--FIRAS limit~\cite{COBE} 
of
$\vert Y \vert < 1.5 \times 10^{-5}$. 
 We conclude\footnote{An additional  contribution to $Y$
  arises as CBR photons pass through 
transitional regions being re-ionized, but is 
two orders of magnitude smaller than the effect we discussed.} that
current observations of the CBR spectrum yield no constraint on the $B=0$ 
universe.

The energy spectrum of uplifted CBR photons shown in
Fig.~\ref{fig:deltaugamma} extends into the visible, falling like
$1/\sqrt{\omega}$. Most of the  energy remaining from nuclear
annihilation resides in this tail. Nevertheless, the diffuse intensity
of the night sky is well above this level. 

\section{The Diffuse Gamma-Ray Spectrum}
\label{sec:diffuse}

In this section,  we use our conservative calculation of the annihilation
rate to determine a lower bound to the CDG signal.  We find that
annihilation in a $B=0$ universe produces far more $\gamma$-rays than are
observed.

The relic spectrum of $\gamma$-rays consists primarily of
photons from $\pi^0$ decay. 
Let $\Phi(E)$ denote the inclusive
photon spectrum in $p \bar{p}$ annihilation, normalized to
$\bar g$, the mean photon
 multiplicity\footnote{The measured photon spectrum
can be found in \cite{Rgamma}\ and is further discussed in
Appendix~\ref{app:electrons}.}.  The average number of photons made per
unit volume, time and energy is $\Phi(E)\,{J/ d}$. These photons scatter
and redshift, leading to a spectral flux of annihilation photons
$F(E,y)$ (number per unit time, area, energy and steradian)
satisfying the transport equation:
\begin{equation}
  \left( y\,{\partial\over \partial y}
    +E\, {\partial\over \partial E} -2 \right)\,F(E,y) =
 -{1\over H(y)}\, 
  \Phi(E)\,{c\, J\over 4\, \pi \, d}+ R(E,y)\;. 
\label{transport}
\end{equation}
The first term on the RHS is the annihilation source
and the second is a scattering sink. We slightly underestimate
$F(E,y)$ by treating
all scattered photons as effectively absorbed. In this case:
\begin{equation}
  R(E,y)={c\,\sigma_\gamma(E)\,n_e(y)\over H(y)}\,F(E,y) \equiv
  g(E,y)\, F(E,y)\;,
\label{eq:sink}
\end{equation}
with $\sigma_\gamma$ the photon interaction cross section and $n_e(y)$
the electron density. For the relevant photon energies, it
matters little whether photons encounter bound or unbound electrons.

Integration of Eqs.~(\ref{transport})--(\ref{eq:sink}) gives the photon
flux today, 
$F(E)\equiv F(E,1)$:
\begin{equation}
  F(E) = \int^{y_R}_{y_S} \frac{c\, J(y')\, \Phi(E y')}{4 \pi\,
    d(y')}  \exp\left[-\int_1^{y'} {dy''\over y''}
    \, g(Ey'',y'')\right] \,
  \frac{dy'}{H(y')\,y^{\prime 3}}\;.
  \label{photnumber}
\end{equation}

\begin{figure}[htb]
  \centering \includegraphics[height=3.5 in]{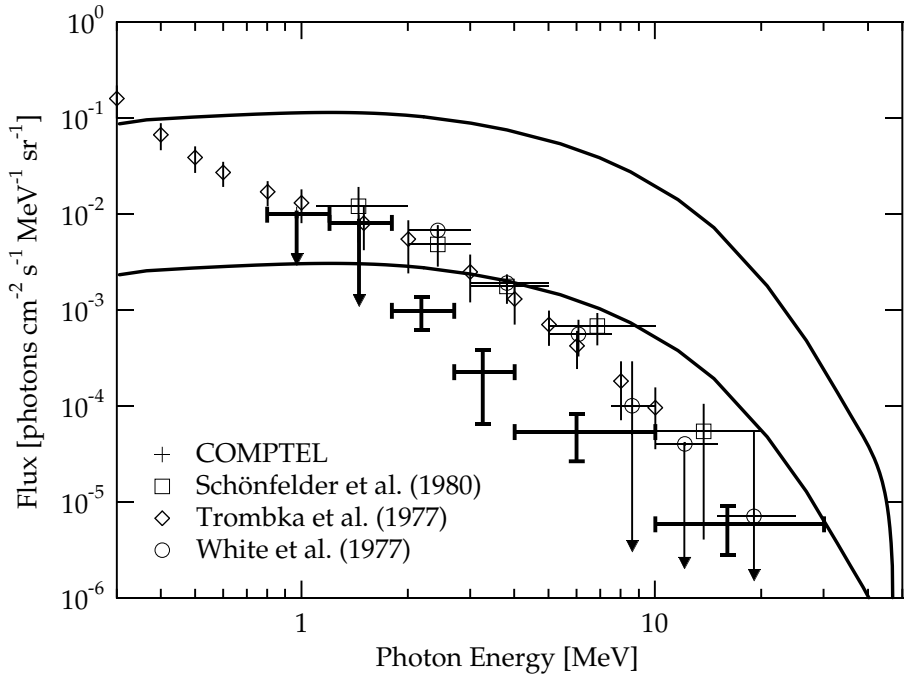}
  \caption{Data \cite{comptel} and expectations for the diffuse
    $\gamma$-ray spectrum.}
  \label{fig:gammas}
\end{figure}

Measurements of the CDG flux are shown in
Fig.~\ref{fig:gammas}.  From 2~MeV to 10~MeV,
preliminary COMPTEL satellite measurements \cite{comptel} lie roughly
an order of magnitude below\footnote{This
discrepancy is attributed by the authors of \cite{comptel}
to a rigidity-dependent background
  correction that the balloon experiments could not perform.} the
earlier balloon data~\cite{gam1}.
Figure~\ref{fig:gammas} also shows our computed signal $F(E)$.
The upper
curve corresponds to the smallest allowed domains,
$d_0=20$ Mpc, the lower curve to
$d_0=1000$ Mpc.  The signal is linear in $1/d_0$. The relic photon
distribution is redshifted from
the production spectrum (which peaks at $E\sim 70$ MeV), and
is slightly depleted at low energies by attenuation. 
 
Our conservative  lower limit to the
$\gamma$-ray signal conflicts with observations  by several orders of
magnitude and over a wide range of energies, 
for all values of $d_0\ltap 10^3$~Mpc, comparable to the size of the universe.
We could argue that the satellite data excludes even larger domain
sizes, but we would soon run into questions of the precise
geometry and location of these nearly horizon-sized domains. 

\section{Closing Loopholes}
\label{sec:cosmo}

Can our `no-go theorem' for the $B=0$ universe be
skirted by changing the  input parameters, modifying our hypotheses, or
including other effects? Here we examine 
the sensitivity of our conclusions to the chosen  values of 
cosmological parameters, to the possible existence of primordial magnetic
fields, and  to the assumed isentropic nature of primordial density
fluctuations.

We used a flat and dark-matter-dominated universe with vanishing
cosmological constant. For this case, the expansion rate is given by  the
simple expression $H(y)=y^{3/2}\, H_0$, with $H_0$ the Hubble constant. 
Other choices for the cosmological parameters  ($\Omega_m\ne 1$
and/or $\Omega_\Lambda\ne 0$) would alter the $y$ dependence of $H(y)$
as follows:
\begin{equation}
\frac{dy}{y} = - H(y)\, dt = -H_0 \left[(1-\Omega)\,y^2 + \Omega_m\, y^3
+\Omega_\Lambda\right]^{1/2} dt \ .
\label{hubble}
\end{equation}
It is only through the modification of $H(y)$ that 
$H_0$, $\Omega_m$ and $\Omega_\Lambda$ affect our results.

We have recomputed the diffuse gamma background (CDG) for a range of
observationally viable values of the cosmological parameters and are
unable to suppress the signal by more than a factor of 2. The reason  is
easily seen. Equation~(\ref{transport})  shows that $J\propto 1/H(y)$, and 
Eq.~(\ref{photnumber}) shows that the CDG flux is proportional to $J/H(y)$,
and hence to $H(y)^{-2}$. To suppress the flux, we must increase $H(y)$
beyond its value at $\Omega_m=1$, $\Omega_\Lambda=0$ and $h=0.75$. No
sensible value of $\Omega_\Lambda$ has much effect at $y\sim 20$, when
most of the CDG flux arises. For $\Omega_m=2$ or $h=0.5$, two
borderline possibilities, the CDG flux would be reduced by about a
factor of two, not altering our conclusions.

We assumed that electrons produced by annihilations travel in
straight lines. This would not be true were there primordial (or
magnetohydrodynamically generated) magnetic fields in the vicinity of
domain boundaries.  
Fields with sufficiently short
correlation lengths and large amplitudes would reduce the electron
range.  If the 
magnetically-reduced range still exceeds $D$, the width of the
depletion zone, the annihilation rate is increased and our conclusions
are strengthened. If the electron range were less than
$D$, electrons would deposit their energy near the
annihilation zone rather than throughout the plasma.  However, 
heating by UV photons alone results in the temperature  profile plotted in
Fig.~\ref{fig:temperature}. Because $J\sim \sqrt{T}$, the CDG
signal cannot be reduced by more than a 
factor of 3 relative to our previous results.
Thus, the existence of magnetic fields at or after recombination
cannot alter our conclusion. 

Finally, we claimed that matter and antimatter domains must touch
by recombination, if they are not to
produce observable (and unobserved) scars in the CBR. Our argument depended
on the absence of strictly isothermal fluctuations at recombination.  
If this hypothesis is false,  matter and
antimatter islands could be separated by regions of vanishing baryon
density, with a uniform  photon distribution throughout. If these 
isothermal voids are so  wide that they persist after recombination,
annihilation might be prevented.  
Annihilation might also be prevented by `wrapping' different
regions with domain walls, whose properties are designed to block the 
penetration of thermal matter while avoiding cosmological
constraints~\cite{walls}. 
We have not further pursued these contrived lines of thought. 

\section{Conclusions}
\label{sec:conc}

Neither the notion of a universe containing islands of antimatter, nor the
exploration of its observable consequences are new. Indeed, the literature
includes diametrically opposed views as to the viability of such models.
The purpose of this paper is to present a class of models
(arguably, the most general) for which the observable universe  
consists of comparable numbers of domains containing either
matter or antimatter. These models are parameterized by the typical domain
size today, $d_0$. Direct searches for annihilation radiation show that
$d_0>20$~Mpc, and future searches for antimatter  among cosmic
rays may increase this lower bound by an order of magnitude.  

We have found constraints on a matter--antimatter universe arising
from phenomena taking place at cosmological distances.  The potentially
observable signals are identified as a distortion of the CBR, and the
production of a relic flux of diffuse gamma-rays (CDG). We have computed
these signals  with conservative assumptions and considerations based  on
empirical evidence, but with  as little theoretical prejudice as possible.
We find that matter--antimatter encounters at domain boundaries are
unavoidable from recombination to the onset of structure
formation.  The detailed dynamics underlying our calculation of the
annihilation rate is complicated. The flow of matter into antimatter (and
vice versa) is diffusive at large $y$ and hydrodynamic at low $y$.
Furthermore, energy deposition by the annihilation debris plays a
crucial role, increasing the annihilation rate by up to two 
orders of magnitude relative to what it would have been if this
effect had been neglected.

Part of the energy released by annihilations at cosmological distances ends
up as microwave photons that would appear as a non-thermal correction to
the cosmic background spectrum. However, we find  that
measurements of the CBR spectrum do not lead to a competitive constraint on
the $B=0$ universe.

High-energy photons produced by annihilations at cosmological distances 
(most of which survive to the current epoch) are redshifted to current
energies of order 1~MeV, thereby contributing  to the diffuse $\gamma$-ray
spectrum.  Our conservative estimate of the relic CDG flux far exceeds its
measured value. Thus, we have ruled out a $B=0$ universe
with domains smaller than
a size comparable to that of the visible universe\footnote{Of course,
  it is not possible to 
exclude the existence of small and distant pockets of
antimatter~\cite{dolsil}.}.  It follows that 
the detection of $Z>1$ antinuclei among cosmic rays would shatter
our current understanding of cosmology, or reveal something
unforeseen in the realm of astrophysical objects.

\bigskip\bigskip\bigskip
\noindent
{\bf Acknowledgements}
\medskip

We would like to thank S.~Ahlen,
M.B.~Gavela, J.~Ostriker, S.~Redner, F.~Stecker and S.C.C.~Ting for
useful conversations. 
This work was supported in part by the Department of Energy under
grant \#DE-FG02-91ER40676 and the National Science Foundation under
grant number NSF-PHYS-92-18167. 

\appendix

\section{The Annihilation Debris}
\label{app:electrons}

Each $p\bar p$ annihilation  produces ${\bar g}\simeq
3.8$ electrons and positrons, and a similar number of photons. 
The photon spectrum has been well measured~\cite{Rgamma} and may be
used to infer the electron spectrum.
The photon distribution peaks at $E_\gamma\simeq 70$ MeV, and the
average photon energy is $\langle 
E_\gamma \rangle \simeq 180$ MeV. The mean pion energy is 
twice that of  the photon.
About $1/4$ of the energy of a charged pion finds its way to an
electron. (The muon retains $\sim\!3/4$ of the charged-pion energy, of
which $\sim\!1/3$ passes to the decay electron.)  Thus, we expect an
electron spectrum peaking at $E_e\sim 35$~MeV with $\langle E_e \rangle
\sim 90$~MeV. 

We must determine various
properties of an annihilation electron in the redshift interval
$20<y<1100$: its mean range, its effect on 
the CBR, the energy it deposits in matter along its trajectory, and
the ionizing effect of its passage.
Three mechanisms control the electrons' motion in the fully ionized
plasma.  With 
$p=\beta\gamma m_e$ and in terms of our fiducial
cosmological parameters, they are:

\begin{itemize}
\item Cosmological redshift:
\begin{eqnarray}
  -\, {dp \over d\,t}\Biggr|_C & = &{\dot R \over R}\, p=H_0 \, p \,y^{3/2}=
K_C(y)\,\beta\gamma\;, \label{dpdtE}\\
K_C(y) & = & 1.3\times 10^{2}\;{y}^{3/2}\  
\frac{\hbox{eV}}{\hbox{Mpc}}\ .
\nonumber
\label{eq:l1}
\end{eqnarray}
\item Collisions with CBR photons:
\begin{eqnarray}
-\, {d\, p \over  d\, t}\Biggr|_\gamma &=&
{4 \, \pi^2 \over 45}\, \sigma_T\,
T_\gamma ^4\, {E^2\over m_e^2}\;\beta\;=
K_\gamma(y)\,\beta\gamma^2\;.
        \label{tal2}\\
  K_\gamma(y) &\simeq& 0.7\; {y}^4\
\frac{\hbox{eV}}{\hbox{Mpc}}\ .
\nonumber
\label{eq:l2}
\end{eqnarray}
\item Collisions with ambient plasma electrons:
\begin{eqnarray}
-\, {d\, p \over d\, t}\Biggr|_M & \simeq &
2\pi \, n_e \, {\alpha^2  \over m_e \beta^2} \;
{\rm ln}\,{\left( {m_e^3\, \beta^2\over 16\,\pi\, n_e\,\alpha} \right)}\;
\simeq
K_M(y)\, {n_e \over n_\infty}\,{1\over \beta^2}
\label{Jac}\\
  K_M(y)&\simeq& 5.5\;{y}^3\ 
\frac{\hbox{eV}}{\hbox{Mpc}}\ .
\nonumber
\label{eq:l3}
\end{eqnarray}
where $n_e$ is the position-dependent electron number density
while $n_\infty$ is its value far enough from a domain boundary to be
unaffected by annihilation and fluid motion.  
\end{itemize}

\subsection{The Range of Annihilation Electrons}
\label{app:range}

Annihilation electrons lose energy as they redshift, but this
mechanism---given by Eq.~(\ref{dpdtE})---is negligible compared with
collisional energy loss throughout the interval
$20<y<1100$. Collisions with CBR photons---given by Eq.~(\ref{tal2}),
for which $dp/dt\propto \gamma^2$---dominate over most of the
trajectory. As an electron becomes non-relativistic, collisions with
background electrons---given by Eq.~(\ref{Jac}), for which
$dp/dt\propto 1/\beta^2$---come into play.  These mechanisms cross
over at $\beta^3\,\gamma^2\simeq 8/y$, a point denoted by $\beta_{eq}$
($\gamma_{eq}$). Some typical values are $\beta_{eq}\simeq 0.62$,
$0.33$, $0.19$ at $y=20$, $200$, $1100$.  

To compute 
the range $L(\gamma_0,\,y)$ of an electron with initial energy $\gamma_0
m_e$, we use Eq.~(\ref{tal2})
throughout its trajectory, and ignore the small effect of
multiple-scattering corrections.  
For $y\gtap 20$ the neglect of other energy-loss mechanisms leads to a
negligible overestimate of $L$.
Integrating Eq.~(\ref{tal2}), we find:
\begin{equation}
L(\gamma_0, y)= {m_e\over K_\gamma }\,\arcsin{\beta_0}
\simeq
0.8\times 10^{-6}\;\left({y_R\over y}\right)^4\ \;{\rm Mpc}\,.
\label{range}
\end{equation}
For an initially relativistic electron $\arcsin{\beta_0}\simeq \pi/2$, 
and the electron range is insensitive to the  initial electron
energy. The  dependence of $L$ on $\gamma_0$ is hereafter suppressed.

The previously established limit on domains of uniform composition is
$d(y)\gtap 20/y$~Mpc.  For $y < 30$, the electron range exceeds this
minimal size and our one-dimensional approximation breaks
down. Because we find a much stronger limit on the minimal domain
size, this complication need not be faced.  The result for the
electron range, including all three sources of energy loss
Eqs.~(\ref{eq:l1})--(\ref{eq:l3}),  is plotted in
Fig.~\ref{fig:sizes}. Throughout the 
relevant redshift interval, $L$ is small compared with the horizon.

\subsection{UV Photons}
\label{app:cbr}

We compute the spectral distortion caused by the passage of one
electron (with initial  energy $E_0=\gamma_0\, m_e$) through a thermal
bath of CBR photons. Compton
scatterings conserve photon number  but
skew the spectrum toward higher energies.
The initial spectral distribution of CBR photons is 
$dn_\gamma/d\omega= (\omega/\pi)^2 {\cal N}(\omega)$, with ${\cal
N}=1/(e^{\omega/T_\gamma}-1)$. 
Let $d^2N(\omega_f,\omega_i)/d\omega_i\,d\omega_f$ denote the
number of photons transferred by one electron  from the frequency interval
$d\omega_i$ to the interval $d\omega_f$. Define:
\begin{equation}
{d^2 N(\omega_f,\omega_i)\over d\,\omega_i\,d\,\omega_f}=
{d^2N(\omega_f,\omega_i)\over d\omega_f\,d\,n}\;
{dn_\gamma(\omega_i)\over d\omega_i}\;.
\label{transfer}
\end{equation}
The function $d^2N(\omega_f,\omega_i)/d\omega_f\,dn$ may be regarded as the
spectral distribution of struck photons of frequency $\omega_f$  produced
during the voyage of one energetic electron through an isotropic,
monochromatic photon gas of unit density and frequency $\omega_i$. 

Let $d\Omega_i(\theta_i,\phi_i)$ be the differential solid
angle about the initial photon direction, and $v_i$ be the relative
speed of the colliding particles. We choose to measure angles
relative to the total momentum direction of the colliding particles.
The function $d^2N/d\omega_f\,dn$ is
obtained by 
averaging the differential transition rate over target photon directions,  
and integrating in time, along the electron trajectory:
\begin{equation}
{d^2N\over d\,\omega_f\,d\,n}=
\int dt \int\frac{d\Omega_i}{4\pi} v_i
\frac{d\sigma}{d\omega_f} \ ,
\label{eq:nshell}
\end{equation}
where we have neglected the small effect of stimulated emission.

The computation is simplified if we note that
$\gamma T_\gamma \ll m_e$, so that the Thompson limit applies and
\begin{eqnarray*}
v_i\frac{d\sigma}{d\omega_f} =
\frac{3\sigma_T}{16 \mu^4 \beta^5 \gamma^{10} \omega_i}\!\!\! 
\,\,\biggl\{ \mu^2\gamma^2\,
(1+2\gamma^2)(1-2\gamma^2 \mu) + (3-4\gamma^2)\mu^4\gamma^4 +4\mu^6\gamma^6\\
  + r\,(r - 2\,\mu\,\gamma^2)\, 
\left[3 - 6\,\mu\,\gamma^2 + \mu^2\gamma^2\,\left(1 + 2\,\gamma^2 \right)  \right]
        \biggr\} \, \Theta(\frac{\mu}{1+\beta} < r <
        \frac{\mu}{1-\beta}) \ ,
\end{eqnarray*}
where $r \equiv \omega_f/\omega_i$ and $\mu\equiv 1-\beta 
\cos\theta_i$.

Carrying out the integrations in Eq.~(\ref{eq:nshell}) gives our
result for $d^2 N/ d\omega_f\,dn$.
(The $dt$ integration is most easily performed by trading $dt$ for $dp$ 
using Eq.~(\ref{tal2}). This integral extends from from $p_0\simeq
m_e\,\gamma_0$ to $p_{eq}\simeq m_e \, \gamma_{eq}\,\beta_{eq}$. The
result is insensitive to the $y$-dependence of $p_f$.)

\subsection{Ionization}
\label{app:ions}

Here we show that the fluid is almost totally ionized by annihilation
electrons at all relevant
times.  The value of the ionization fraction, $x$, results from a
compromise between the recombination and ionization rates.
Annihilation
electrons ionize the material they traverse both directly, via
electron-atom collisions as described by Eq.(\ref{Jac}), or
indirectly, via the UV showers discussed in Appendix~\ref{app:cbr}.
We discuss the latter effect, which is more important.

Many of the photons up-scattered by annihilation electrons have energies
exceeding the hydrogen binding energy ($B=13.6$~eV), and can ionize
hydrogen atoms via $\gamma+H\rightarrow e+p$. 
The photoionization cross section for hydrogen atoms
in their ground state,
$\sigma_K$,  falls rapidly from a very large threshold value
$\sigma_K(B) \simeq 8\times10^{-18}$~cm$^2$:
\begin{equation}
\sigma_K(\omega)\simeq  \sigma_K(B)\,(B/\omega)^3\,\Theta(\omega-B)\ .
\label{sigk}\end{equation}
We compute the effective ionization cross section $\bar\sigma_K$ for
the entire UV shower associated with a single electron by integrating
the product of 
$\sigma_K$ with the photon number distribution:
\begin{equation}
\bar{\sigma}_K \equiv \int {d^2N\over d\omega_i\,d\omega_f}
\,\sigma_K(\omega_f) \,d\omega_f\,d\omega_i\simeq 1.4\times10^{-13}\;{\rm cm}^2\;
\sqrt{1100/y}\;.
\label{krate}
\end{equation}
This cross section is four orders of magnitude larger than
$\sigma_K(B)$ and reflects the large number of photons scattered by a single
electron.

The total ionization rate is the difference of the photonionization
rate and the recombination rate.
The former is obtained by multiplying the effective
ionization cross section for a single annihilation electron
$\bar\sigma_K$ by the 
flux of electrons. Because half of the  $e^\pm$ produced in an
annihilation zone move to either  side, the flux is half the multiplicity
${\bar g}$ times the annihilation rate $J$.
The total ionization rate $\dot x$ (per second and
per baryon) is: 
\begin{equation}
\dot{x}  =  
\frac{\bar g}{2}\, J\, (1-x)\,  \bar{\sigma}_K - n_e \, x^2 \,
\langle \sigma_{\rm rec}\; v_e \rangle \label{recomb} \;,
\label{UVion1}
\end{equation}
where the recombination coefficient to all states but the ground state 
is
\begin{equation}
\langle \sigma_{\rm rec}\; v_e \rangle \simeq 1.14\times 10^{-13}\; T^{-1/2}
\left[1-2.20\, \log T + 0.814\; T^{1/3}\right] \;
\hbox{cm}^3\, \hbox{s}^{-1} \ .
\end{equation}
The
coefficient of $1-x$ in Eq.~(\ref{UVion1}) is 
much greater than the coefficient
of $x^2$ at all relevant epochs.  Consequently, the ionization is 
very close to one:
\begin{equation}
1-x \simeq  \frac{2\, n_e \langle \sigma_{\rm rec}\; v_e \rangle}{{\bar g}\,
  J\, \bar{\sigma}_K} \ll 1\; .
\label{eq:fullion}
\end{equation} 
In the previous argument, no allowance
was made for photon absorption despite the large photoionization
cross section. Because the (quasi-)equilibrium ionization is
nearly total,  UV photons are unlikely to encounter
atoms.

Near the region of electron production, the UV photon shower has not fully
developed, so that the ionization is smaller than
Eq.~(\ref{eq:fullion}) indicates.  Our calculation of $dN/d\omega_f dn$ can
be modified to treat this case.  We find that the UV flux near 
the annihilation zone is sufficient to maintain total ionization to
within a few percent.

The UV flux generated by annihilation is sufficient to prevent recombination
by producing and sustaining almost total ionization. However, for large
values of $d_0$, regions lying  far from  domain boundaries 
recombine as in a standard cosmology. A
moving front develops between  
ionized and recombined regions as the UV flux progresses. The velocity
of the front is \[v_f \sim {c\over 
1+\xi}\;,\] where $\xi$ is the ratio of the nucleon number density to that
of the incident UV flux. We find $v_f\sim c/3$ at $y=1100$ and $v_f\sim
c$ at $y=20$. The intense energy deposition taking place within the front
makes an unobservably small  contribution to the Sunyaev--Zeldovich parameter.

\subsection{Energy Deposition}
\label{app:deposit}

We compute the heat function $H_\epsilon$: the energy deposited in
the plasma, per unit volume and time, by annihilation electrons and UV
photons.

In regions within the electron range, this function is dominated by 
the primary electron contribution.
For
$\gamma>\gamma_{eq}$, collisions with CBR photons determine the evolution
of the electron velocity according to Eq.~(\ref{tal2}).
Denoting the energy deposition to matter for this portion of the
trajectory by ${\cal E}_1$, we integrate Eq.~(\ref{Jac}) to find:
\begin{equation}
{\cal E}_1 = K_M(y)\,{n\over n_\infty}\,
\int_0^{L'}\,{dx\over\beta^2} = {K_M\over K_\gamma}\;{m_e\,c^2\over
\beta_{eq}\gamma_{eq}}\,{n\over n_\infty}\ .
\label{elos1}
\end{equation}
Here $L'$ is the distance traveled when $\gamma = \gamma_{eq}$:
\begin{equation}
L'(y)=(m_e/K_\gamma) \arccos \beta_{eq}\ .
\label{lprime}
\end{equation}

Most of the remaining energy, ${\cal
  E}_2 \simeq (\gamma_{eq} -1)\; m_ec^2$, is deposited in matter over a
relatively small distance interval. 
About one third of the energy deposition to matter takes place during this
short stopping stage. In the following analysis, we ignore this term,
thereby underestimating electron heating
by $\sim$~30\%, and slightly underestimating  the production
of CDG photons.

Electrons arise as an isotropic flux from the thin annihilation
zone of width $A$. The
angular average, per electron, of the energy deposition to matter at a
distance $l\gg A$ from this zone is:
\begin{equation}
-\Biggl\langle{dE\over dl}\Biggr\rangle_M\simeq\int_{l+A}^{L'}
{dx\over x}\; \biggl[K_M(y)\,{n\over n_\infty}\,{1\over \beta^2}
\biggr] \;,
\label{dedl}
\end{equation}
where 
the integration variable is the distance traveled by an electron along
its trajectory. 
Within the depletion zone  $\langle{dE/dl}\rangle_M$ is a
slowly-varying function of $l$ that is roughly proportional to the
electron density $n_e$: 
\begin{equation}
\biggl\langle{dE\over dl}\biggr\rangle_M \simeq a\, \ y^3\
\frac{\hbox{eV}}{\hbox{Mpc}}\ ,
\end{equation} 
where $10 \ltap a \ltap 20$. For our computations we use the smallest
value of $a$.  

Half of the  $e^\pm$ produced in an annihilation zone move to either side.
Thus the $e^\pm$ flux is ${\bar g}\, J/2$, and the electron
contribution to the heat
function is:
\begin{equation}
H_\epsilon=-\, \frac{\bar g}{2}\,J\, \Biggl\langle{d\,E\over
  dl}\Biggr\rangle_M\ .
\label{Heat}
\end{equation}
The UV photon contribution is small in comparison with that of the electrons.

Outside the electron range, only
UV photons contribute to $H_\epsilon$. In an ionizing collision,
$\gamma+H\rightarrow e+p$, 
the mean kinetic energy $\delta E$ of the recoiling photoelectron
is:
\begin{equation}
\delta\,E = \frac{1}{\bar\sigma_K}\int \omega\,{d^2N\over d\omega\,d\omega_i}
\,\sigma_K(\omega)\, d\omega\,d\omega_i - B \simeq 5.4\ \hbox{eV}\;,
\label{eq:recoil}
\end{equation}
where the cross sections and distribution function are those of
Appendix~\ref{app:cbr}. The rate per unit volume of such collisions is 
$J {\bar \sigma}_K (1-x) n_e {\bar g}/2 $.
Using Eq.~(\ref{eq:fullion}) we express this rate as
$n_e^2 \langle \sigma_{\rm rec} v_e \rangle$.
Multiplying by the mean recoil energy, we obtain the heat function:
\begin{equation}
H_\epsilon = \delta E\, \langle \sigma_{\rm rec} v_e \rangle \,n_e^2\;.
\end{equation}
The UV photon flux has disappeared from this
expression, reflecting the quasi-equilibrium state of the ionization.
As a welcome consequence, $H_\epsilon$ is insensitive to
additional UV photons arising from annihilation zones other than the
nearest.

\end{document}